\documentclass[aps,prl,twocolumn,superscriptaddress]{revtex4}
\usepackage{graphicx}
\usepackage{amssymb}
\usepackage{amsmath}
\usepackage{graphicx}
\usepackage{epsfig}
\usepackage{color}
\usepackage{bm}

\begin{document}

\title{A Simple Solvable Model for Heavy Fermion Superconductivity from the Two-Fluid Normal State}
\author{Jiangfan Wang}
\email[]{jfwang@hznu.edu.cn}
\affiliation{School of Physics, Hangzhou Normal University, Hangzhou, Zhejiang 311121, China}
\author{Yu Li}
\affiliation{Institute of Quantum Materials and Physics, Henan Academy of Sciences, Zhengzhou, Henan 450046, China}
\affiliation{Kavli Institute for Theoretical Sciences, University of Chinese Academy of Sciences, Beijing 100190, China}
\author{Yi-feng Yang}
\email[]{yifeng@iphy.ac.cn}
\affiliation{Beijing National Laboratory for Condensed Matter Physics and Institute of Physics,
Chinese Academy of Sciences, Beijing 100190, China}
\affiliation{School of Physical Sciences, University of Chinese Academy of Sciences, Beijing 100190, China}
\affiliation{Songshan Lake Materials Laboratory, Dongguan, Guangdong 523808, China}
\date{\today}

\begin{abstract}
We propose an exactly solvable momentum-space Kondo-BCS model to study heavy fermion superconductivity. The Kondo interaction is local in momentum space, which can be derived from an Anderson lattice with a Hatsugai-Kohmoto interaction between $f$-electrons. By increasing the Kondo interaction, the model exhibits a crossover from a weak-coupling BCS superconductor to a strong-coupling heavy fermion superconductor featured with a large gap ratio and a large specific heat jump anomaly. Accordingly, the normal state evolves from a Fermi liquid above the BCS superconductor to a non-Fermi liquid two-fluid state above the heavy fermion superconductor. The two-fluid normal state also leads to two types of Cooper pairs, one between conduction electrons, the other  between composite fermions formed by conduction electrons and $f$-spins, which is responsible for the strong coupling behaviors of heavy fermion superconductivity.
\end{abstract}

\maketitle

\textit{Introduction.}---Heavy fermion compounds are  prototypical  strongly correlated electron systems that exhibit unconventional superconductivity \cite{Pfleiderer2009, White2015, Steglich2016}. Due to strong electron correlations and interplay between multiple degrees of freedom, heavy fermion superconductors (HFSCs) are known for their rich diversity of order parameter structures and  pairing mechanisms \cite{Lonzarich1998-CePd2Si2,Petrovic2001-CeCoIn5,Yuan2003-CeCu2Si2,Park2006-CeRhIn5,Schuberth2016-YbRh2Si2,Aoki2019, Ran2019,Holmes2004,Ramshaw2015}. Despite of such a diversity, they share important macroscopic features. For example, they are often formed out of an unusual two-fluid normal state, where physical quantities are contributed by two different parts: a Kondo liquid part associated with the itinerant heavy electrons, and a classical spin liquid part associated with the residual unhybridized local spins \cite{Nakatsuji2004, YangPRL2008, YangNature2008, YangPNAS2012, YangReview2016}.  The two-fluid  normal state is also responsible for the coexistence of superconductivity and magnetic order observed in many heavy fermion materials \cite{Thompson2012,YangPNAS2014}. Other common features of HFSCs include their low transition temperatures, large jump anomaly of the specific heat, and often a large ratio between the superconducting gap and the transition temperature  \cite{White2015,Stewart2017,Kohori2001,Mito2001,Zheng2001,Mizukami2011,Daghero2012}. Therefore, seeking an efficient way to describe both the two-fluid normal state and the universal features of HFSC emerged from it is an important issue in this area.

Previous studies of HFSCs are either based on phenomenological models with effective pairing interaction \cite{Nishiyama2013,YuLi2018}, or Anderson/Kondo lattice models that inevitably rely on analytical or numerical approximations \cite{Xavier2008,Flint2010,Liu2012,Wu2015,Hu2021}. Most of these studies focus on certain microscopic properties, but fail to make a connection between the superconductivity and the two-fluid normal state. The problem roots in the difficulty of dealing with the strong $f$-electron interaction and the lack of an microscopic explanation for the two-fluid behaviors. Recently, it is found that the exactly solvable Hatsugai-Kohmoto (HK) model may provide a convenient tool to study strongly correlated problems \cite{HK1992, Phillips2020,Phillips2022,Phillips2023,YinZhong2022,YuLi2022}. The model assumes an all-to-all nonlocal interaction that transforms to a local one in momentum space, which represents a stable interacting fixed point describing the Mott physics \cite{Phillips2023}. Inspired by this, we recently  proposed a  momentum-space Kondo lattice model that allows for a microscopic derivation of the two-fluid phenomenology \cite{Wang2023}. The key point is the often neglected  nonlocal Kondo effect in heavy fermion systems, which causes a partial screening of $f$-spins in momentum space and leads to two-fluid behaviors.

In this paper, we introduce an exactly solvable momentum-space Kondo-BCS model, and study superconductivity emerged from the two-fluid normal state. By increasing the Kondo coupling, we found a crossover from a BCS superconductor to a  HFSC as the corresponding normal state evolves from a Fermi liquid state to a non-Fermi liquid (NFL) two-fluid state. The HFSC is featured with a significant enhancement of the gap ratio  and the specific heat jump anomaly relative to the BCS values, indicating its  strong coupling nature. In addition, we found two types of Cooper pairs in the superconducting phase, one formed by conduction electrons, the other formed by the composite fermions (composite objects of conduction electrons and $f$-spins), which is responsible for the strong coupling properties of HFSC. Our model provides a useful tool to study both the unconventional superconductivity and normal state in heavy fermion systems.

\textit{Model.}---We begin with the $\mathbf{k}$-space Kondo lattice model:
\begin{eqnarray}
H_K&=&\sum_{\mathbf{k}\alpha}\epsilon_\mathbf{k}c_{\mathbf{k}\alpha}^\dagger c_{\mathbf{k}\alpha}+J_K\sum_{\mathbf{k}}\boldsymbol{s}_{\bf k}\cdot \boldsymbol{S}_{\mathbf{k}},
\label{eq:H_K}
\end{eqnarray}
where $c_{\mathbf{k}\alpha}^\dagger $ creates a conduction electron with  dispersion $\epsilon_\mathbf{k}$, $\boldsymbol{s}_{\bf k}=\frac{1}{2}\sum_{\alpha\beta}c_{\mathbf{k}\alpha}^\dagger \boldsymbol{\sigma}_{\alpha\beta}c_{\mathbf{k}\beta}$ and  $\boldsymbol{S}_\mathbf{k}$ are the spins of the conduction electron and $f$-electron defined at each momentum $\mathbf{k}$, $J_K$ is the Kondo coupling.

$H_K$ can be derived from a Schrieffer-Wolff (SW) transformation of the Hatsugai-Kohmoto-Anderson lattice model where the $f$-electrons are described by the HK Hamiltonian \cite{YinZhong2022}:
\begin{eqnarray} H_f=\sum_{\mathbf{k}}\xi_\mathbf{k}n_\mathbf{k}^f+U\sum_{\mathbf{k}} n_{\mathbf{k}\uparrow}^f n_{\mathbf{k}\downarrow}^f.\label{eq:Hf}
\end{eqnarray}
Here an extra kinetic term for $f$-electrons is added, and $n_{\bf k}^f$ ($n_{\bf k \sigma}^f$) is the occupation number (per spin). The HK interaction in $H_f$ leads to a breakdown of Fermi liquid and gives rise to Mott physics \cite{Phillips2022,Phillips2023}. At half-filling, the exact solution of $H_f$ reveals a quantum phase transition between a NFL metal and a Mott insulator as $U$ increases \cite{Phillips2020}. This leads to a well defined $\mathbf{k}$-space  $f$-spin operator $\boldsymbol{S}_\mathbf{k}=\frac{1}{2}\sum_{\alpha\beta}f_{\mathbf{k}\alpha}^\dagger \boldsymbol{\sigma}_{\alpha\beta}f_{\mathbf{k}\beta}$ with $n_{\bf k}^f=1$ deep inside the Mott insulating state. By including the hybridization with conduction electrons,
\begin{equation} H_c+H_{hyb}=\sum_{\mathbf{k}\alpha}\epsilon_\mathbf{k}c_{\mathbf{k}\alpha}^\dagger c_{\mathbf{k}\alpha}+\mathcal{V}\sum_{\mathbf{k}\alpha}\left(c_{\mathbf{k}\alpha}^\dagger f_{\mathbf{k}\alpha}+H.c.\right),
\end{equation}
and performing a SW transformation to project out the $f$-electron charge fluctuations, one obtains Eq. (\ref{eq:H_K}) with $J_K\approx  8|\mathcal{V}|^2/U$ \cite{supp}. The resulting $\mathbf{k}$-space Kondo interaction represents a nonlocal scattering between $c$ and $f$ electrons. Previous studies suggest that such nonlocal Kondo interaction indeed plays an important role in heavy fermion systems \cite{Wang2021_nonlocal,Wang2022_Z2,Wang2022_3IK}.

\begin{figure}
	\centering\includegraphics[scale=0.57]{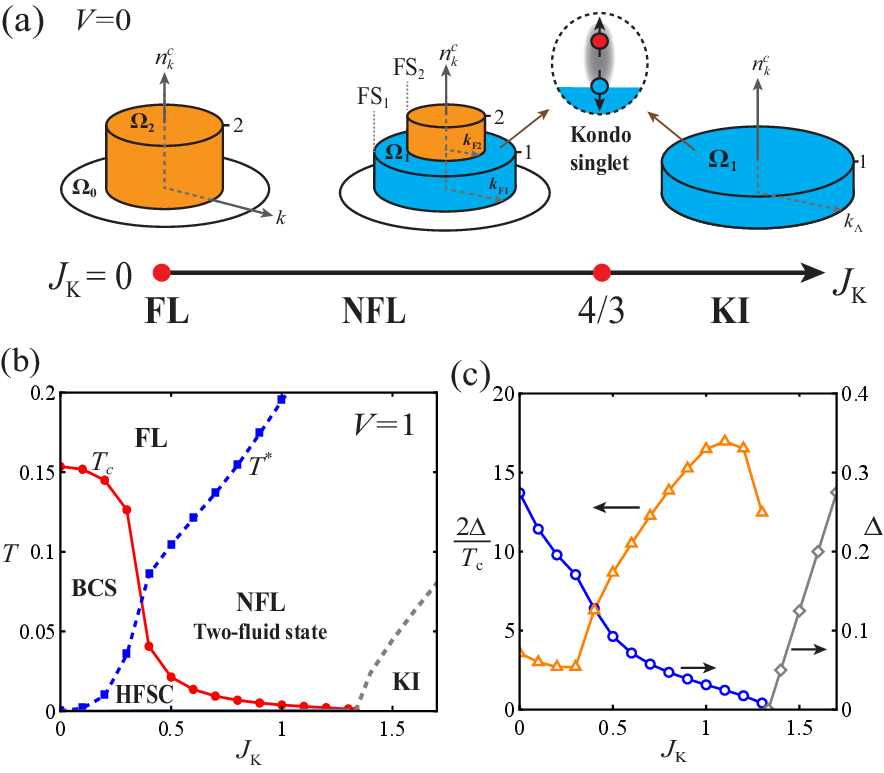}
	\caption{(a) Three different ground states of the $\mathbf{k}$-space Kondo lattice model $H_K$ as $J_K$ increases: Fermi liquid (FL), non-Fermi liquid (NFL) metal and Kondo insulator (KI). The conduction electron occupation number $n_{\bf{k}}^c$ is shown for each phase. Kondo singlets only reside on the singly occupied region $\Omega_1$.  (b) The phase diagram of the $\mathbf{k}$-space Kondo-BCS model with pairing interaction $V=1$. (c) The energy gap $\Delta$ for the superconducting (blue) and KI (gray) states, and the superconducting gap ratio $2\Delta/T_c$ (orange) at different $J_K$.}
	\label{Fig1}
\end{figure}

\textit{Normal states.}---Figure (\ref{Fig1}a) shows the ground state phase diagram  of $H_K$ and the corresponding conduction electron occupation number $n_{\bf k}^c$.  For simplicity, here we have assumed a parabolic dispersion $\epsilon_\mathbf{k}=k^2/2\pi-1$ with an ultraviolet cutoff $k_\Lambda=2\sqrt{\pi}$, so that the half-band-width $D=1$ serves as the energy unit, and the electron density is $n_c=\mathcal{N}_s^{-1} \sum_{\bf k}n_{\bf k}^c=1$ throughout our calculations. At $J_K=0$, the conduction electrons form a Fermi liquid (FL) that is completely decoupled from the $f$-spins. A finite $J_K$ destroys the Fermi liquid by replacing the original Fermi surface with a singly occupied momentum region $\Omega_1$,  separated from the empty region ($\Omega_0$) and  the doubly occupied region ($\Omega_2$)  by two filling surfaces at momenta $k_{F1}$ and $k_{F2}$, here denoted as $\text{FS}_1$ and $\text{FS}_2$. The appearance of three occupation regions with two filling surfaces is a hallmark of NFL ground states in HK-like models \cite{HK1992,Baskaran1991,TKNg2020}. The $f$-spins form $\mathbf{k}$-space Kondo singlets with conduction electrons only in the $\Omega_1$ region, while remain free in  $\Omega_0$ and $\Omega_2$, since the Pauli principle forbids spin-flip scattering at a doubly occupied $\mathbf{k}$ point. As $J_K$ increases, the $\Omega_1$ region also enlarges, until it covers the entire momentum space at $J_K=4/3$, beyond which the system enters into a  Kondo insulating (KI) phase due to the complete screening of $f$-spins.

The partial Kondo screening in the NFL state has two consequences: 1) The Mott insulated $f$-electrons in $\Omega_1$ become itinerant through the Kondo hybridization, such that the total number of charge carriers per spin is now counted by the ``Fermi volume'' enclosed by $\text{FS}_1$, which is larger than the Fermi volume at $J_K=0$, but smaller than the standard large Fermi volume of a heavy Fermi liquid state. Such a partially enlarged Fermi volume is also found in previous large-$N$ \cite{Wang2021_nonlocal,Wang2022_Z2,Wang2022_3IK} or numerical calculations \cite{JiaLin2023} for strongly frustrated or one-dimensional Kondo lattices where nonlocal Kondo interactions play an important role. 2) The itinerant Kondo singlets in $\Omega_1$ form a Kondo liquid, with an ``order parameter'' satisfying a universal scaling consistent with the phenomenological two-fluid model \cite{Wang2023}. The remaining unhybridized $f$-spins in $\Omega_0$ and $\Omega_2$ form another fluid that is generally referred to as a ``spin liquid''. Moreover, the conduction electrons in $\Omega_0$ and $\Omega_2$ also form a third liquid, as implicitly assumed in the two-fluid model \cite{YangReview2016}. The above NFL properties persist within a finite temperature region below the characteristic Kondo coherence temperature $T^*$ as shown in Fig. (\ref{Fig1}b), above which the thermal fluctuations destroy the Kondo singlets and the normal state becomes a Fermi liquid controlled by the trivial fixed point at $J_K=0$.

\textit{Superconductivity.}---To study the superconducting instability, we consider the simplest $s$-wave pairing interaction between conduction electrons,
\begin{equation} H_V=-\frac{V}{\mathcal{N}_s}\sum_{\mathbf{k}\mathbf{k'}}c_{\mathbf{k}\uparrow}^\dagger c_{\mathbf{-k}\downarrow}^\dagger c_{-\mathbf{k'}\downarrow} c_{\mathbf{k'}\uparrow},
\end{equation}
and calculate the electron pair-binding energy, $E_b=\langle \psi|H'|\psi \rangle-\langle G|H'|G \rangle$, where  $H'=H_K+H_V$, $|G\rangle$ is the ground state of $H_K$, and $|\psi\rangle$ is the state with an additional Cooper pair \cite{supp}. The resulting $E_b$ satisfies
\begin{equation}
	1=\frac{V}{16D}\ln \left|\frac{(2D-E_b)^4(3J_K-E_b)}{-(3J_K/2-E_b)^4 E_b}\right|, \label{eq:CI}
\end{equation}
and the numerical results are shown in the Supplementary Materials. At $J_K=0$, Eq. (\ref{eq:CI}) reduces to the BCS result.  Increasing $J_K$ reduces the absolute value $|E_b|$, but $E_b$ stays negative throughout the entire NFL state for arbitrarily small $V$, indicating Cooper instability. Next, we choose a finite $V$ and perform a BCS mean-field decomposition of $H_V$, and solve the combined Hamiltonian:
\begin{eqnarray}
	H&=&H_K+H_\text{BCS}, 	\label{eq:H} \\
	H_\text{BCS}&=&\Delta_c\sum_{\mathbf{k}} c_{\mathbf{k}\uparrow}^\dagger c_{\mathbf{-k}\downarrow}^\dagger +H.c.+\frac{\mathcal{N}_s \Delta_c^2}{V}, \notag
\end{eqnarray}
where $\Delta_c=-V\mathcal{N}_s^{-1}\sum_{\mathbf{k}}\left\langle c_{\mathbf{-k},\downarrow} c_{\mathbf{k}\uparrow} \right\rangle$ is the pairing amplitude. Eq. (\ref{eq:H}) is exactly solvable, since it can be written as $H=\frac{1}{2}\sum_{\mathbf{k}}H_\mathbf{k}$, where each $H_\mathbf{k}$ is a conserved quantity with a $64$-dimensional Hilbert space, hence can be exactly diagonalized.

The phase diagram for $V=1$ is shown in Fig. (\ref{Fig1}b). A superconducting phase is found for  $J_K<4/3$, with the transition temperature $T_c$ decreasing monotonically with increasing $J_K$. A rapid suppression of $T_c$ is found between $J_K=0.3$ and $0.4$, where the phase boundary changes its sign of curvature. At this point, the normal state also evolves from a FL to a NFL two-fluid state. The crossover temperature $T^*$ is determined by a broad maximum of the specific heat coefficient associated with the Kondo effect. Without pairing interaction, $T^*$ decreases almost linearly with $J_K$ and vanishes at $J_K=0$. With finite pairing interaction, $T^*$ continuous to exist below $T_c$ albeit being suppressed, which separates the superconducting phase into two regions: a weak coupling BCS superconductor and a strong coupling HFSC. The superconducting transition is continuous for $J_K\leq 0.5$ but becomes weakly first-order for $J_K\geq 0.6$, where $\Delta_c$ jumps discontinuously at $T_c$ \cite{supp}. We notice that first-order transition is not rare in studies of superconductivity emerging from non-Fermi liquids \cite{YuLi2022,Patel2018}. The interaction driven superconductor-to-insulator quantum phase transition at $J_K=4/3$ is another interesting topic that may have experimental relevance \cite{Sun2022,Sun2023}, which we leave for future investigation.

Fig. (\ref{Fig1}c) plots the zero temperature energy gap $\Delta$  and the gap ratio $2\Delta/T_c$. Note that $\Delta$ is determined by the electron spectral function, which is generally unequal to the pairing amplitude $\Delta_c$, except for $J_K=0$ where the BCS relation holds. The gap ratio is quite close to (less than) the universal BCS value 3.53 for $J_K\leq 0.3$, but becomes significantly enhanced for $J_K\geq 0.4$, with a maximal value $2\Delta/T_c\approx 17$ at $J_K=1.1$. Such an extremely large gap ratio is also found  in quantum critical models where the pairing glue has a power-law local susceptibility \cite{THLee2018,Chubukov2019}, or superconductivity emerged from incoherent metals with Sachdev-Ye-Kitaev interactions \cite{Patel2018}. In our case, we attribute the large gap ratio at $J_K\geq 0.4$ to the unusual Kondo liquid in the normal state, which suppresses $T_c$ and induces another type of Cooper pair that is strongly-coupled in nature.

\begin{figure}
	\centering\includegraphics[scale=0.66]{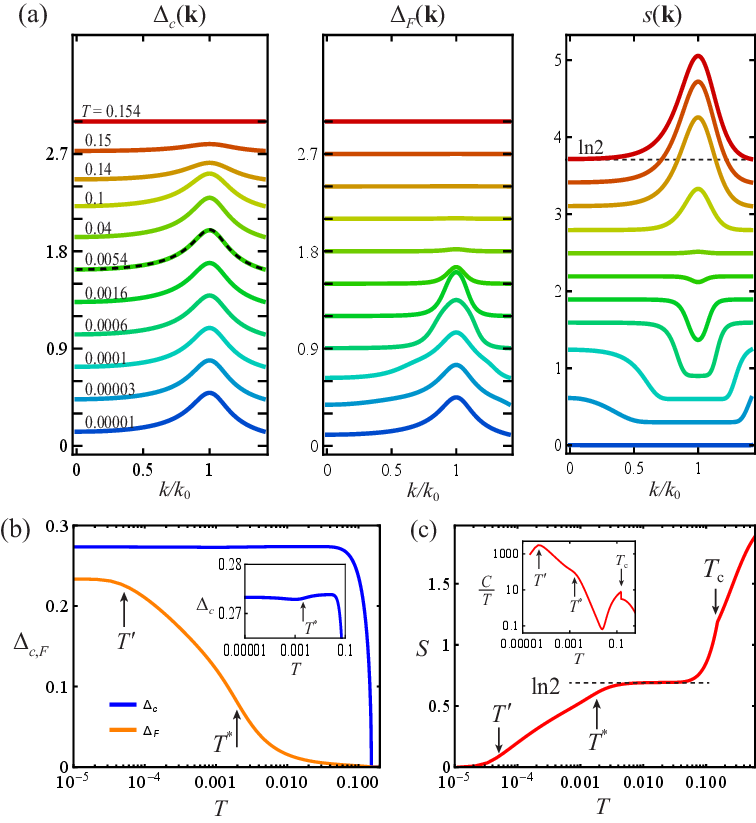}
	\caption{(a) Evolution of $\Delta_c(\mathbf{k})$, $\Delta_F(\mathbf{k})$ and $s(\mathbf{k})$ with decreasing temperature at $J_K=0.1$. For clarity, the data curves are shifted with each other by a constant 0.3.  The black dashed line in $\Delta_c(\mathbf{k})$  is the BCS result, $\Delta_c(\mathbf{k})=\Delta_c/(2\sqrt{\epsilon_\mathbf{k}^2+\Delta_c^2})$. (b) The pairing amplitudes $\Delta_c$ and $\Delta_F$ as functions of temperature at $J_K=0.1$. The inset shows a zoom-in view of $\Delta_c$. (c) The entropy as a function of temperature at $J_K=0.1$. The inset shows the specific heat coefficient. }
	\label{Fig2}
\end{figure}

To distinguish the two regions of the superconducting state, we define a composite fermion operator $F_{\mathbf{k}\alpha}$, and study its pairing correlation:
\begin{equation}
	\Delta_{F}(\mathbf{k})=-\langle F_{\mathbf{-k}\downarrow}F_{\mathbf{k}\uparrow}\rangle,\qquad 	F_{\mathbf{k}\alpha}=\sum_\beta \boldsymbol{\sigma}_{\alpha\beta}\cdot \boldsymbol{S}_\mathbf{k} c_{\mathbf{k}\beta}. \label{eq:CF}
\end{equation}
In heavy fermion literature, the composite fermion is usually defined in the coordinate space, which transforms to a convolution in the momentum space \cite{Coleman2009,Assaad2021,Coleman2023}. For the $\mathbf{k}$-space Kondo model studied here, Eq. (\ref{eq:CF}) is a more appropriate definition, since the SW transformation used to derive $H_K$ also transforms the $f$-electron operator $f_{\mathbf{k}\alpha}$ to $F_{\mathbf{k}\alpha}$ \cite{supp}. Therefore, one can roughly view $F_{\mathbf{k}\alpha}$ as the renormalized $f$-electrons in the Kondo lattice model.  A finite $\Delta_F(\mathbf{k})$ indicates presence of composite fermion Cooper pairs, which requires both pairing interaction and Kondo entanglement.

\begin{figure}
	\centering\includegraphics[scale=0.66]{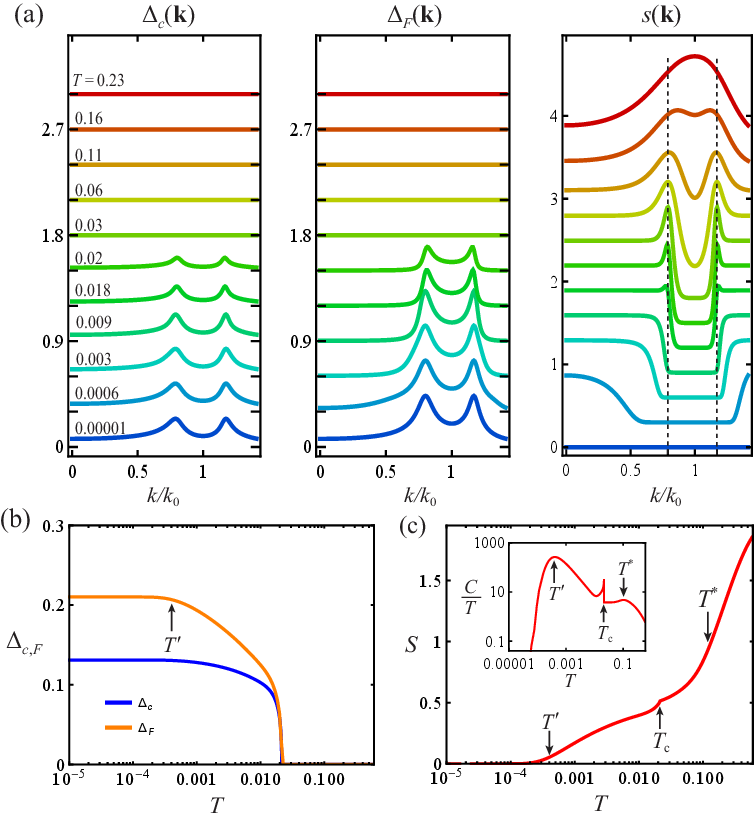}
	\caption{(a) Temperature evolution of $\Delta_c(\mathbf{k})$, $\Delta_F(\mathbf{k})$ and $s(\mathbf{k})$ for $J_K=0.5$. The data curves are shifted with each other by a constant 0.3.  The dashed lines in $s(\mathbf{k})$ mark the positions of the two filling surfaces at $k_{F1}=1.17k_0$ and $k_{F2}=0.79k_0$. (b) The pairing amplitudes $\Delta_c$ and $\Delta_F$ as functions of temperature at $J_K=0.5$. (c) The entropy as a function of temperature at $J_K=0.5$. The inset shows the specific heat coefficient. }
	\label{Fig3}
\end{figure}

Fig. (\ref{Fig2}a) compares the temperature evolution of  $\Delta_c(\mathbf{k})=-\langle c_{\mathbf{-k}\downarrow}c_{\mathbf{k}\uparrow}\rangle$, $\Delta_F(\mathbf{k})$, and the entropy distribution $s(\mathbf{k})=-\frac{1}{2}\text{Tr}[\rho_\mathbf{k}\ln \rho_\mathbf{k}]$ ($\rho_\mathbf{k}$ is the density matrix of $H_{\mathbf{k}}$) for $J_K=0.1$. Above $T_c\approx 0.152$, both $\Delta_c(\mathbf{k})$ and $\Delta_F(\mathbf{k})$ are zero, while $s(\mathbf{k})$ shows a peak around the bare ($J_K=0$) Fermi momentum $k_0=\sqrt{2\pi}$,  with a constant background $\ln 2$ contributed by the free $f$-spins. Below $T_c$, $\Delta_c(\mathbf{k})$ becomes finite for all values of $\mathbf{k}$, while $\Delta_F(\mathbf{k})$ is almost zero within a broad range of temperature. The shape of $\Delta_c(\mathbf{k})$ can be fitted perfectly by the BCS formula $\Delta_c(\mathbf{k})=\Delta_c/(2\sqrt{\epsilon_\mathbf{k}^2+\Delta_c^2})$ as shown in Fig. (\ref{Fig2}a).  The entropy peak around $k_0$ is gradually consumed by the conduction electrons forming Cooper pairs, leaving a constant $\ln2$ plateau of local spins, which is also clearly seen in the integrated entropy $S=\mathcal{N}_s^{-1}\sum_\mathbf{k}s(\mathbf{k})$ as shown in Fig. (\ref{Fig2}c). Below the characteristic temperature $T^*\approx 0.002$, a large amount of  $f$-spins start to combine with conduction electrons to form composite fermion Cooper pairs, as indicated by the peak of $\Delta_F(\mathbf{k})$ and the dip of $s(\mathbf{k})$  around $k_0$.  Interestingly, the rapid development of composite pairs is accompanied by a slight suppression of the conduction electron pair amplitude, as shown by the temperature dependence of  $\Delta_c=V\mathcal{N}_s^{-1}\sum_\mathbf{k}\Delta_c(\mathbf{k})$ and $\Delta_F=V\mathcal{N}_s^{-1}\sum_\mathbf{k}\Delta_F(\mathbf{k})$ in  Fig. (\ref{Fig2}b). This suggests that some   Cooper pairs unbind themselves in order to form the composite pairs, indicating that these are indeed different types of Cooper pairs.

Just below $T^*$, the peak of $\Delta_F(\mathbf{k})$ is concentrated in the vicinity of $k_0$, which quickly extends to the entire momentum space as the system approaches zero temperature. The same happens to the dip (gap) of $s(\mathbf{k})$, indicating all the $f$-spins finally become a part of the superconducting condensate. This process happens within a small temperature window, but consumes the extensive $f$-spin entropy in the $\Omega_0$ and $\Omega_2$ regions, giving rise to a huge peak of $C/T$ at another characteristic temperature $T'$. Due to this huge peak, the broad maximum of $C/T$ at $T^*$ now appears as a shoulder, as shown in the inset of Fig. (\ref{Fig2}c). Below $T'$, $\Delta_F$  saturates to a constant, while $S$ approaches zero. This characteristic temperature exists for all $0<J_K<4/3$, see for example Fig. (\ref{Fig3}) for $J_K=0.5$. However it may not occur in real materials, since a more natural way to consume the extensive spin entropy is to form a nearby or coexisting magnetic order as observed in many HFSCs.

\begin{figure}[b]
	\centering\includegraphics[scale=0.78]{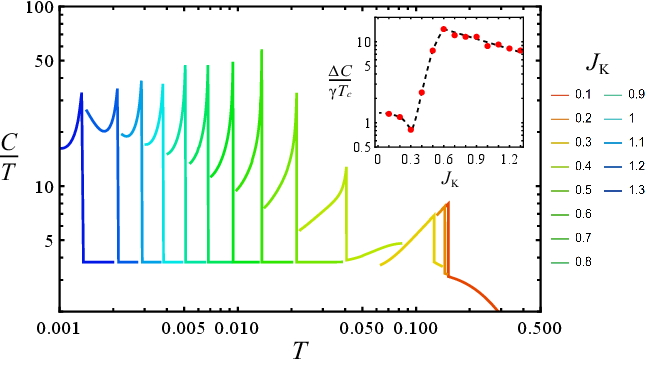}
	\caption{The specific heat coefficient around the superconducting transition temperature at different $J_K$. Inset: the specific heat anomaly $\Delta C/\gamma T_c$ as a function of $J_K$.}
	\label{Fig4}
\end{figure}

Fig. (\ref{Fig3}) shows the same physical quantities at $J_K=0.5$. Above $T_c\approx 0.021$, both $\Delta_c(\mathbf{k})$ and $\Delta_F(\mathbf{k})$ are zero, while $s(\mathbf{k})$ evolves from a single peak to two peaks centered around the two filling surfaces at  $k_{F1}=1.17k_0$ and $k_{F2}=0.79k_0$. The entropy depletion between $k_{F1}$ and $k_{F2}$ is due to the formation of Kondo singlets in the $\Omega_1$ region. It  causes a broad maximum of $C/T$ at $T^*\approx 0.1$, as shown in the inset of Fig. (\ref{Fig3}c). Superconductivity occurs at $T_c\approx 0.021$, above which the Kondo singlets in $\Omega_1$ region have already been fully developed. Therefore, both the conduction electrons and composite fermions form Cooper pairs immediately below $T_c$. The two-peak structure of $\Delta_c(\mathbf{k})$ and $\Delta_F(\mathbf{k})$ suggests that the Cooper pairs are mainly formed by quasiparticles around the two filling surfaces, similar to the HK model \cite{Phillips2020,YuLi2022}. Below $T_c$, the pairing amplitudes $\Delta_c$ and $\Delta_F$ increase monotonically with decreasing temperature as shown in Fig. (\ref{Fig3}b). Different from the case of $J_K=0.1$, here $\Delta_F$ is much larger than $\Delta_c$, indicating that the composite fermion Cooper pairs play a dominant role.

Since the superconducting transitions for $J_K\geq 0.4$ require $f$-spins around the two filling surfaces to form composite Cooper pairs, they consume more entropy and lead to large specific heat anomaly $\Delta C/\gamma T_c$, as shown in Fig. (\ref{Fig4}). Here, the Sommerfeld coefficient $\gamma=3.77$ is universal for the two-fluid normal state within $0<J_K<4/3$, which is only slightly larger than the non-interacting value $\gamma=\pi^2/3\approx 3.29$ at $J_K=0$. It may require local-in-space Kondo interaction to obtain a $\gamma$ as large as in real heavy fermion materials. As $J_K$ increases, $\Delta C/\gamma T_c$ first decreases for $J_K\leq 0.3$, then increases rapidly for $0.4\leq J_K\leq 0.6$, and decreases again for $J_K\geq 0.6$ where the superconducting transition becomes first-order. For continuous transitions within $0<J_K<0.6$, the evolution of $\Delta C/\gamma T_c$ follows that of the gap ratio, indicating the same microscopic origin behind them. For a rough comparison, experiments on two of the most studied strong coupling HFSCs, CeCoIn$_5$ and CeRhIn$_5$, show $\Delta C/\gamma T_c= 4.5\sim 4.7$ \cite{Petrovic2001-CeCoIn5,Ikeda2001},  $2\Delta/T_c=6 \sim 10$ \cite{Park2008,Kohori2001}, and $\Delta C/\gamma T_c=4.2$ \cite{Mito2001},  $2\Delta/T_c=5$ \cite{Park2009}, consistent with our results at $J_K=0.4\sim 0.5$.

\textit{Discussion.}---We have checked different values of $V$ and the results remain qualitatively the same. Our method can be conveniently generalized to $d$-wave or $p$-wave pairing interactions. In all cases, the composite fermion Cooper pairs will inevitably arise due to the combined effect of Kondo and pairing interactions, and lead to important universal features of HFSC. An implication of our study is that both conduction electron Cooper pairs and composite fermion Cooper pairs may exist in real HFSCs consistent with their two-fluid normal states, and the latter is expected to play a dominant role. What remains unexplored in this work is the microscopic mechanism behind the pairing interaction, which is often associated with the $f$-spin fluctuations in real materials.  It is  found that a Heisenberg exchange interaction between $\boldsymbol{S}_{\bf k}$ and $\boldsymbol{S}_{-\mathbf{k}}$ indeed induces pairing correlation between conduction electrons, which may require additional scattering to establish phase coherence \cite{Wang2023}. In general, the $\mathbf{k}$-space Kondo model has the advantage over conventional real space models that it can be solved exactly, while at the same time captures important universal features of heavy fermion materials, thus opens a new window towards the ultimate solving of heavy fermion problems.

The authors thank Y. Zhong for stimulating discussion. J.W. and Y.-F.Y. are supported by the National Natural Science Foundation of China (Grants No. 12174429, No. 11974397), the National Key Research and Development Program of China (Grant No. 2022YFA1402203), and the Strategic Priority Research Program of the Chinese Academy of Sciences (Grant No. XDB33010100). And Y.L. is supported by the Fundamental Research Funds for the Central Universities (Grant No. E2E44305).

\end{document}


\title{A Simple Solvable Model for Heavy Fermion Superconductivity from the Two-Fluid Normal State\\
\vspace{0.2cm}
- Supplemental Material -}
\author{Jiangfan Wang}
\email[]{jfwang@hznu.edu.cn}
\affiliation{School of Physics, Hangzhou Normal University, Hangzhou, Zhejiang 311121, China}
\author{Yu Li}
\affiliation{Institute of Quantum Materials and Physics, Henan Academy of Sciences, Zhengzhou, Henan 450046, China}
\affiliation{Kavli Institute for Theoretical Sciences, University of Chinese Academy of Sciences, Beijing 100190, China}
\author{Yi-feng Yang}
\email[]{yifeng@iphy.ac.cn}
\affiliation{Beijing National Laboratory for Condensed Matter Physics and Institute of Physics, 
	Chinese Academy of Sciences, Beijing 100190, China}
\affiliation{School of Physical Sciences, University of Chinese Academy of Sciences, Beijing 100190, China}
\affiliation{Songshan Lake Materials Laboratory, Dongguan, Guangdong 523808, China}
\date{\today}

\maketitle

In this supplementary material, we provide some important derivations in detail and additional figures that may be helpful for readers.

\subsection{I. Derivation of the $\mathbf{k}$-space Kondo model}

In this section, we derive the $\mathbf{k}$-space Kondo model from the Hatsugai-Kohmoto-Anderson lattice model \cite{YinZhong2022}:
\begin{eqnarray}
	H&=&H_c+H_f+H_{hyb} \notag \\
	H_c&=&\sum_{\mathbf{k}\alpha}\epsilon_\mathbf{k}c_{\mathbf{k}\alpha}^\dagger c_{\mathbf{k}\alpha} \notag \\
	H_f&=&\sum_{\mathbf{k}}\xi_\mathbf{k}n_\mathbf{k}^f+U\sum_{\mathbf{k}} n_{\mathbf{k}\uparrow}^f n_{\mathbf{k}\downarrow}^f, \notag \\
	H_{hyb}&=&\mathcal{V}\sum_{\mathbf{k}\alpha}\left(c_{\mathbf{k}\alpha}^\dagger f_{\mathbf{k}\alpha}+H.c.\right). \label{eq:S1}
\end{eqnarray} 
Here $\xi_{\bf k}=\epsilon_{\bf k}^f-\mu_f$, $\epsilon_{\bf k}^f\in [-D_f,D_f]$ is the $f$-electron dispersion, $\mu_f$ is its chemical potential. At $\mathcal{V}=0$, the ground state of $H_f$ is known exactly \cite{Phillips2020}. At half-filling ($\mu_f=U/2$), the ground state is a NFL metal for $U<2D_f$, and a Mott insulator for $U>2D_f$. To describe heavy fermion systems, here we assume half-filling and $U\gg 2D_f$ .

The $\mathbf{k}$-space Kondo model can be derived from a Schrieffer-Wolf (SW) transformation of the above model. We start with the low energy states of $H_c+H_f$: 
\begin{equation}
	|l\rangle:\quad|\Psi\rangle_c \otimes \prod_{\bf k}| n_{\bf k}^f=1\rangle,  \label{eq:S2}
\end{equation}
where $|\Psi\rangle_c$ denotes the eigenstates of $H_c$. Turning on the hybridization $\mathcal{V}$ will induce charge fluctuations and lead to high energy states with $n_{\bf q}^f\neq 1$ at some momentum $\mathbf{q}$:
\begin{eqnarray}
	|h\rangle:&& c_{\mathbf{q}\sigma}|\Psi\rangle_c \otimes f_{\mathbf{q}\sigma}^\dagger\prod_{\bf k}| n_{\bf k}^f=1\rangle,  \notag \\
	 && c_{\mathbf{q}\sigma}^\dagger|\Psi\rangle_c \otimes f_{\mathbf{q}\sigma}\prod_{\bf k}| n_{\bf k}^f=1\rangle. \label{eq:S3}
\end{eqnarray}
Here we have ignored high energy states with more than one empty or doubly occupied momentum points, which only  give higher order ($O(\mathcal{V}^4)$) corrections. Using Eq. (\ref{eq:S2}) and (\ref{eq:S3}) as a basis, one can rewrite Eq. (\ref{eq:S1}) into the following matrix form separating the low (L) and high energy (H) subspaces:
\begin{equation}
	H=\begin{bmatrix}
		H_\text{L} &  M^\dagger \\
		 M & H_\text{H} 
	\end{bmatrix}, 
\label{eq:S4}
\end{equation}
where $(H_\text{L})_{ll'}=\langle l|H_c+H_f |l'\rangle$, $(H_\text{H})_{hh'}=\langle h|H_c+H_f |h'\rangle$, and $M_{hl}=\langle h|H_{hyb} |l\rangle$ mixes different subspaces. The SW transformation is a canonical transformation $\mathcal{U}=e^S$ that turns Eq. (\ref{eq:S4}) into block-diagonal form:
\begin{equation}
	\mathcal{U}\begin{bmatrix}
		H_\text{L} &  M^\dagger \\
		  M & H_\text{H} 
	\end{bmatrix} \mathcal{U}^\dagger = \begin{bmatrix}
	    H^* & 0 \\
	    0 & H' 
    \end{bmatrix}. \label{eq:S5}
\end{equation}
Then  $H_\text{eff}=P_\text{L}H^*P_\text{L}$ provides the effective low-energy Hamiltonian, where $P_\text{L}=\sum_l |l\rangle \langle l|$ is the projection operator. Eq. (\ref{eq:S5}) is satisfied up to order $O(\mathcal{V}^2)$ if 
\begin{equation}
	S= \begin{bmatrix}
		0 & -s^\dagger \\
		s & 0 
	\end{bmatrix},\quad s_{hl}=\frac{M_{hl}}{E_h^\text{H}-E_l^\text{L}}, \label{eq:S6}
\end{equation} 
where $E_h^\text{H}$ and $E_l^\text{L}$ are the eigenvalues of $H_\text{H}$ and $H_\text{L}$. Substituting Eq. (\ref{eq:S6}) into Eq. (\ref{eq:S5}) leads to $H^*=H_L+\Delta H$, with  
\begin{eqnarray}
(\Delta H)_{ll'}&=&-\frac{1}{2}\sum_{h}M_{lh}^\dagger M_{hl'}\left(\frac{1}{E_h^\text{H}-E_l^\text{L}}+\frac{1}{E_h^\text{H}-E_{l'}^\text{L}}\right) \notag \\
 &=&\Delta H_{ll'}^{f^1+e^-\leftrightarrow f^2}+\Delta H_{ll'}^{f^1\leftrightarrow f^0+e^-} ,
\end{eqnarray}
where the last line corresponds to two different types of charge fluctuations in Eq. (\ref{eq:S3}). Using 
\begin{equation}
E_h^\text{H}-E_l^\text{L}=\left\{	\begin{array}{cc}
	\xi_{\bf k}+U-\epsilon_{\bf k}, & f^1+e^-\leftrightarrow f^2 \\
	-\xi_{\bf k}+\epsilon_{\bf k}, & f^1\leftrightarrow f^0+e^-
	\end{array}\right. ,
\end{equation} 
one has
\begin{equation}
	H_{ll'}^{f^1+e^-\leftrightarrow f^2}=-\mathcal{V}^2\sum_{\mathbf{k}\alpha\beta}\frac{\langle l|c_{\mathbf{k}\alpha}^\dagger f_{\mathbf{k}\alpha}f_{\mathbf{k}\beta}^\dagger c_{\mathbf{k}\beta}|l'\rangle }{\xi_{\bf k}+U-\epsilon_{\bf k}}
	\label{eq:S9}
\end{equation}
and 
\begin{equation}
H_{ll'}^{f^1\leftrightarrow f^0+e^-}=-\mathcal{V}^2\sum_{\mathbf{k}\alpha\beta}\frac{\langle l|f_{\mathbf{k}\beta}^\dagger c_{\mathbf{k}\beta} c_{\mathbf{k}\alpha}^\dagger f_{\mathbf{k}\alpha}|l'\rangle}{-\xi_{\bf k}+\epsilon_{\bf k}}. \label{eq:S10}
\end{equation}
One important observation from Eq. (\ref{eq:S9}) and (\ref{eq:S10}) is that the $\mathbf{k}$ point must be singly occupied by  conduction electrons in both $|l\rangle$ and $|l'\rangle$ in order to give a nonzero matrix element for $\alpha\neq \beta$, which then leads to the  Kondo spin-flip scattering. Therefore, the $\mathbf{k}$-space Kondo interaction only operates in the singly occupied ($\Omega_1$) region, as is indeed revealed by the exact solution of $H_K$. After rearranging the four fermion operators in Eq. (\ref{eq:S9}) and (\ref{eq:S10}), we obtain
\begin{eqnarray}
	P_\text{L}\Delta H P_\text{L}=\sum_{\mathbf{k} }J_K(\mathbf{ k})\boldsymbol{s}_{\bf k}\cdot \boldsymbol{S}_{\bf k}.
\end{eqnarray} 
with the Kondo coupling
\begin{eqnarray}
	J_K(\mathbf{k})&=&2\mathcal{V}^2\left(\frac{1}{\xi_{\bf k}+U-\epsilon_{\bf k}}+\frac{1}{-\xi_{\bf k}+\epsilon_{\bf k}}\right)\notag \\
	&\approx&\frac{8\mathcal{V}^2}{U},
\end{eqnarray}
where we have used $\xi_{\bf k}=\epsilon_{\bf k}^f-U/2$ and $U\gg D, D_f$. Since $P_{\text{L}}H_\text{L} P_\text{L}=H_c+ \text{constant}$, we finally obtain the $\mathbf{k}$-space Kondo model $H_K$ described in the main article.

\subsection{II. Cooper instability}

To study the pair-binding energy of conduction electrons, we consider the following variational wave function:
\begin{equation}
	|\psi\rangle=\sum_{\mathbf{k}\in\Omega_0}\alpha_{\bf k}b_{\bf k}^\dagger |G\rangle+\sum_{\mathbf{k}\in\Omega_1}\beta_{\bf k}b_{\bf k}^\dagger |G\rangle,
\end{equation}
where $b_{\bf k}^\dagger=c_{\mathbf{k}\uparrow}^\dagger c_{-\mathbf{k}\downarrow}^\dagger$ creates a Cooper pair, $\alpha_{\bf k}$ and $\beta_{\bf k}$ are two variational parameters, and 
\begin{eqnarray}
	|G\rangle=\prod_{\bf k}|g\rangle_{\bf k}
\end{eqnarray} 
with
\begin{eqnarray}
	|g\rangle_{\bf k}=\left\{\begin{array}{cc}
	\frac{1}{\sqrt{2}}(|0\Uparrow\rangle+|0\Downarrow\rangle), & {\bf k}\in \Omega_0 \\
	\frac{1}{\sqrt{2}}(|\uparrow\Downarrow\rangle-|\downarrow \Uparrow\rangle), & {\bf k}\in \Omega_1\\
	\frac{1}{\sqrt{2}}(|2\Downarrow\rangle+|2 \Uparrow\rangle), & {\bf k}\in \Omega_2
    \end{array}\right. 
\end{eqnarray}
is the ground state of $H_K$. Here $|\phi \sigma\rangle$ denotes  the conduction electron state $\phi=0, \uparrow, \downarrow, 2$ and the $f$-spin state $\sigma=\Uparrow,\Downarrow$ at each momentum point. Note that each $f$-spin in $\Omega_0$ and $\Omega_2$ has a two-fold spin degeneracy, while here we choose the unpolarized state to calculate the binding energy \cite{YuLi2022}. The normalization condition of $|\psi \rangle$ gives the constraint 
\begin{eqnarray}
	1=\langle \psi |\psi \rangle=\sum_{\mathbf{k}\in\Omega_0}|\alpha_{\bf k}|^2+\frac{1}{4}\sum_{\mathbf{k}\in\Omega_1}|\beta_{\bf k}|^2. 
	\label{eq:con}
\end{eqnarray}
The binding energy is defined as 
\begin{eqnarray}
	E_b=\langle \psi |H'|\psi\rangle-\langle G|H'|G\rangle,
\end{eqnarray}
where $H'=H_K+H_V$ is the total Hamiltonian including the pairing interaction 
\begin{equation} H_V=-\frac{V}{\mathcal{N}_s}\sum_{\mathbf{k}\mathbf{k'}}c_{\mathbf{k}\uparrow}^\dagger c_{\mathbf{-k}\downarrow}^\dagger c_{-\mathbf{k'}\downarrow} c_{\mathbf{k'}\uparrow}. 
\end{equation}
The final result of $E_b$ is 
\begin{eqnarray}
	E_b&=&2\sum_{\mathbf{k}\in\Omega_0}|\alpha_{\bf k}|^2 \epsilon_{\bf k}+\sum_{\mathbf{k}\in\Omega_1}\left(\frac{\epsilon_{\bf k}}{2}+\frac{3J_K}{8}\right)|\beta_{\bf k}|^2 \notag\\
	&&-\frac{V}{\mathcal{N}_s}\sum_{\mathbf{k,k'}\in\Omega_0}\alpha_{\bf k'}^*\alpha_{\bf k}-\frac{V}{16\mathcal{N}_s}\sum_{\mathbf{k,k'}\in\Omega_1}\beta_{\bf k'}^*\beta_{\bf k} \notag\\
	&&-\frac{V}{4\mathcal{N}_s}\sum_{\mathbf{k}\in\Omega_0,\mathbf{k'}\in\Omega_1}\left(\alpha_{\bf k}\beta_{\bf k'}^*+\alpha_{\bf k}^*\beta_{\bf k'}\right).
\end{eqnarray}

\begin{figure}[b]
	\centering\includegraphics[scale=0.7]{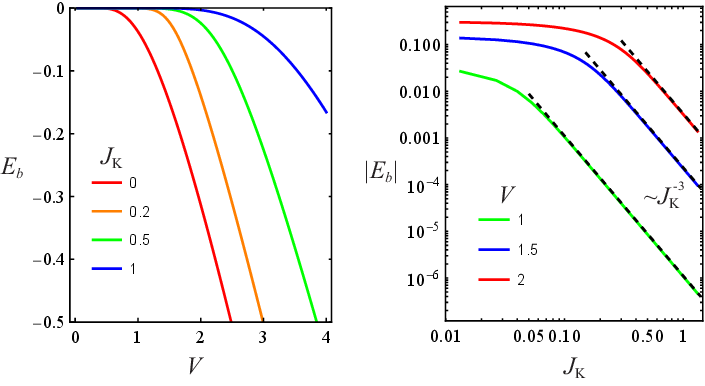}
	\caption{The Cooper pair binding energy $E_b$ as a function of $V$ (left panel) and $J_K$ (right panel), respectively. The dashed lines show $J_K^{-3}$ dependence of $|E_b|$ at large $J_K$. }
	\label{figS1}
\end{figure}

We then use the method of Lagrange multiplier to find the minimum of $E_b$. By taking partial derivatives of the function $Q=E_b-\lambda (\langle \psi |\psi \rangle-1)$ with respect to $\alpha_{\bf k}^*$ and $\beta_{\bf k}^*$, we have
\begin{eqnarray}
	0&=&(2\epsilon_{\bf k}-\lambda)\alpha_{\bf k}-y \notag\\
	0&=&\left(2\epsilon_{\bf k}-\lambda+\frac{3J_K}{2}\right)\beta_{\bf k}-y
\end{eqnarray} 
where $y=\frac{V}{\mathcal{N}_s}\left(\sum_{\mathbf{k}\in \Omega_0}\alpha_{\bf k}+\frac{1}{4}\sum_{\mathbf{k}\in\Omega_1}\beta_{\bf k}\right)$. Multiplying the above equations by $\alpha_{\bf k}^*$ and $\frac{1}{4}\beta_{\bf k}^*$ respectively and summing over the momentum leads to $\lambda=E_b$. Thus we have
\begin{eqnarray}
	\alpha_{\bf k}&=&\frac{y}{2\epsilon_{\bf k}-E_b}, \notag\\
	\beta_{\bf k}&=&\frac{y}{2\epsilon_{\bf k}+\frac{3J_K}{2}-E_b}.
\end{eqnarray}
Substituting the above result into Eq. (\ref{eq:con}) leads to 
\begin{eqnarray}
	1&=&\frac{1}{\mathcal{N}_s}\sum_{\mathbf{k}\in \Omega_0}\frac{V}{2\epsilon_{\bf k}-E_b} \notag \\
	&&+\frac{1}{4\mathcal{N}_s}\sum_{\mathbf{k}\in \Omega_1}\frac{V}{2\epsilon_{\bf k}+\frac{3J_K}{2}-E_b}.
\end{eqnarray}
Using the density of states 
\begin{equation}
	\rho(\omega)=\frac{1}{\mathcal{N}_s}\sum_{\bf k}\delta(\omega -\epsilon_{\bf k})=\frac{1}{2D},
\end{equation}
we finally obtain
\begin{equation}
	1=\frac{V}{16D}\ln \left|\frac{(2D-E_b)^4(3J_K-E_b)}{-(3J_K/2-E_b)^4 E_b}\right|.
\end{equation}
For $J_K=0$ and small $V$, the above equation reduces to the BCS result $E_b=-2D e^{-4D/V}$. For $J_K,D\gg E_b$, we obtain the asymptotic behavior $E_b\propto -D^4 J_K^{-3}e^{-16D/V}$, as shown in Fig. (\ref{figS1}).

\subsection{III. First-order transitions at large $J_K$}

\begin{figure}
	\centering\includegraphics[scale=0.95]{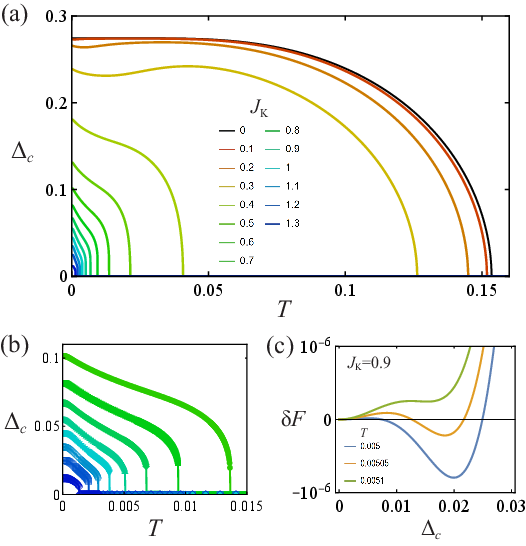}
	\caption{(a) The mean-field order parameter $\Delta_c$ as functions of temperature at different $J_K$ and $V=1$. (b) An enlarged view of $\Delta_c$ for $0.6\leq J_K\leq 1.3$, showing discontinuous jumps at $T_c$. (c) The free energy difference $\delta F=F(\Delta_c)-F(0)$ as functions of $\Delta_c$ around $T_c$ for $J_K=0.9$. }
	\label{figS2}
\end{figure}

Here we provide the full data of the temperature dependence of mean-field order parameter $\Delta_c$ at different $J_K$ for $V=1$, as shown in Fig. (\ref{figS2}a). The general behavior of $\Delta_c$ for $J_K\leq 0.3$ is similar to the BCS result at $J_K=0$, except for the low temperature minima at which $\Delta_F$ increases rapidly. For $J_K\geq 0.4$, $\Delta_c$ has a large slope at intermediate temperatures, which is quite different from the BCS theory. For $J_K\geq 0.6$, $\Delta_c$ jumps discontinuously at $T_c$ as shown in Fig. (\ref{figS2}b), indicating first-order transitions. Accordingly, the free energy as a function of $\Delta_c$ exhibits two minima at $\Delta_c=0$ and $\Delta_c\neq 0$ within a small temperature window (see Fig. (\ref{figS2}c) for $J_K=0.9$), corresponding to the metastable normal and superconducting states in the coexisting region of a weakly first-order transition. 

\subsection{IV. Derivation of the Composite fermion}

The fact that the composite fermion correspond to the SW transformation of $f$-electron has been noted earlier in Ref. \cite{Assaad2021}. Here we derive this result for the $\mathbf{k}$-space Kondo model. We first notice that the $f$-electron annihilation (creation) operator itself connects the low-energy and high-energy subspaces, so it appears in the off-diagonal block of the matrix form in Eq. (\ref{eq:S4}). The canonical transformation $\mathcal{U}=e^S\approx 1+S$ then transforms the $f$ annihilation operator into 
\begin{eqnarray}
	&&e^S \begin{bmatrix}
		  & f_{\mathbf{k}\sigma} \\
		  f_{\mathbf{k}\sigma} & 
	\end{bmatrix}e^{-S}\\
&=&\begin{bmatrix}
	-s^\dagger f_{\mathbf{k}\sigma}-f_{\mathbf{k}\sigma}s & f_{\mathbf{k}\sigma} \\
	f_{\mathbf{k}\sigma} & sf_{\mathbf{k}\sigma}+f_{\mathbf{k}\sigma}s^\dagger 
\end{bmatrix}+O(\mathcal{V}^2). \notag
\end{eqnarray}   
Using the expression of $s$ in Eq. (\ref{eq:S6}), we have the following renormalized $f$-electron operator in the low-energy subspace:
\begin{eqnarray}
	&&-P_\text{L}\left(s^\dagger f_{\mathbf{k}\sigma}+f_{\mathbf{k}\sigma}s\right)P_\text{L}\notag \\
	&=&-\sum_{ll'h}\frac{|l\rangle\langle l|\mathcal{V}\sum_{\mathbf{p}\alpha}c_{\mathbf{p}\alpha}^\dagger f_{\mathbf{p}\alpha}+H.c. |h\rangle \langle h|f_{\mathbf{k}\sigma}|l'\rangle \langle l'|}{E_h^\text{H}-E_l^\text{L}}\notag \\
	&&-\sum_{ll'h}\frac{|l\rangle \langle l|f_{\mathbf{k}\sigma}|h\rangle \langle h|\mathcal{V}\sum_{\mathbf{p}\alpha}c_{\mathbf{p}\alpha}^\dagger f_{\mathbf{p}\alpha}+H.c. |l'\rangle  \langle l'|}{E_h^\text{H}-E_{l'}^\text{L}}\notag \\
	&\approx&-\frac{2\mathcal{V}}{U}P_\text{L}\sum_{\alpha}\left(f_{\mathbf{k}\alpha}^\dagger c_{\mathbf{k}\alpha}f_{\mathbf{k}\sigma}+f_{\mathbf{k}\sigma}f_{\mathbf{k}\alpha}^\dagger c_{\mathbf{k}\alpha}\right)P_\text{L}\notag\\
	&=&\frac{4\mathcal{V}}{U}F_{\mathbf{k}\sigma}
\end{eqnarray}
where 
\begin{eqnarray}
F_{\mathbf{k}\sigma}&=&-\frac{1}{2}P_\text{L}\sum_\alpha	\left(f_{\mathbf{k}\alpha}^\dagger c_{\mathbf{k}\alpha}f_{\mathbf{k}\sigma}+f_{\mathbf{k}\sigma}f_{\mathbf{k}\alpha}^\dagger c_{\mathbf{k}\alpha}\right)P_\text{L}\notag\\
&=&P_\text{L}\left(f_{\mathbf{k},-\sigma}^\dagger f_{\mathbf{k}\sigma}c_{\mathbf{k},-\sigma}+f_{\mathbf{k}\sigma}^\dagger f_{\mathbf{k}\sigma}c_{\mathbf{k}\sigma}-\frac{1}{2}c_{\mathbf{k}\sigma}\right)P_\text{L}\notag\\
&=&S_{\bf k}^{-\sigma}c_{\mathbf{k},-\sigma}+\frac{1}{2}\left(f_{\mathbf{k}\sigma}^\dagger f_{\mathbf{k}\sigma}-f_{\mathbf{k},-\sigma}^\dagger f_{\mathbf{k},-\sigma}\right)c_{\mathbf{k}\sigma}\notag\\
&=&S_{\bf k}^{-\sigma}c_{\mathbf{k},-\sigma}+\text{sgn}(\sigma)S_{\bf k}^z c_{\mathbf{k}\sigma} \notag \\
&=&\sum_\beta \boldsymbol{\sigma}_{\sigma\beta}\cdot \boldsymbol{S}_{\bf k}c_{\mathbf{k}\beta}
\end{eqnarray}
is just the composite fermion operator defined in the main article.